\documentclass[showpacs,aps,prd]{revtex4}
\usepackage{amsmath}
\usepackage{mathrsfs}
\usepackage{amssymb}
\input epsf

\begin{document}
\title{Improved analysis on the semi-leptonic decay $\Lambda_c\rightarrow\Lambda\ell^+\nu$
from QCD light-cone sum rules}
\author{Yong-Lu Liu, Ming-Qiu Huang and Dao-Wei Wang}
\affiliation{Department of Physics, National University of Defense Technology, Hunan 410073, China}
\date{\today}
\begin{abstract}
With the renewed distribution amplitudes of $\Lambda$, we present a
reanalysis on the semi-leptonic decay
$\Lambda_c\rightarrow\Lambda\ell^+\nu$ by use of the light-cone sum
rule approach with two kinds of interpolating currents. The form
factors describing the decay process are obtained and used to
predict the decay width. With the inclusion of up to twist-$6$
contributions the calculations give the decay width
$\Gamma=(10.04\pm0.88)\times10^{-14}\mbox{GeV}$ for
Chernyak-Zhitnitsky-type(CZ-type) current and
$\Gamma=(6.45\pm1.06)\times10^{-14}\mbox{GeV}$ for Ioffe-type
current. The Ioffe-type interpolating current is found to be better
for the estimation of the decay rate from a comparison with
experimental data.
\end{abstract}
 \maketitle
\section{Introduction}
\label{sec1} Flavor changing decays of heavy hadrons are of great
interest in the heavy flavor physics due to their ability to provide
useful information on the quark structure of the hadrons and reveal
the nature of the weak interactions. In particular, the decay of the
$c$-quark baryons can give us various charm related
Cabibbo-Kobayashi-Maskawa matrix elements, which are the main
ingredients of the standard model. Furthermore, a thorough
understanding of the standard model itself needs a comprehension of
the flavor changing dynamics. However, such a comprehension is
difficult contemporarily since form factors characterizing those
processes are nonperturbative quantities that need to be determined
by some nonperturbative method. This paper aims to give a
preliminary determination of the form factors of the exclusive
semi-leptonic decay $\Lambda_c\rightarrow\Lambda\ell^+\nu$. In the
calculation we will use the method of QCD sum rules on the light
cone \cite{BBK}, which in the past has been successfully applied to
various problems in heavy meson physics, see \cite{lcsr} for a
review.

The light-cone sum rule (LCSR) is a non-perturbative method
developed from the standard technique of the traditional QCD sum
rules from Shifman, Vainshtein, and Zakhavov(SVZ sum rules)
\cite{svzsum}, which comes as the remedy for the conventional
approach in which vacuum condensates carry no momentum \cite{ball}.
The main difference between SVZ sum rule and LCSR is that the
short-distance Wilson OPE(Operator Product Expansion) in increasing
dimension is replaced by the light-cone expansion in terms of
distribution amplitudes (DAs) of increasing twist, which were
originally used in the description of the hard exclusive process
\cite{HEP}. In recent years there have been many applications of
LCSR to baryons. The nucleon electromagnetic form factors were
studied for the first time in Refs. \cite{nucleon1,nucleon2} and
later in Refs. \cite{nucleon3,nucleon4,Aliev} for a further
consideration. Several nucleon related processes gave fruitful
results within LCSR, the weak decay $\Lambda_b\to p\ell\nu_\ell$ was
considered in both full QCD and HQET LCSR \cite{lbtop}. The
generalization to the $N\gamma\Delta$ transition form factor was
worked out in Ref. \cite{ntodelta}. We have given the applications
of LCSR on other $J^P=\frac{1}{2}^+$ octet baryons in Refs.
\cite{DAs,EMff}.

In this paper we will make use of the LCSR approach to study the decay process $\Lambda_c\rightarrow\Lambda\ell^+\nu$, which has been preliminarily studied in the
previous work \cite{Wang}. The improvement of the present paper is to use the renewed distribution amplitudes provided in Ref. \cite{DAs}, and adopt two different
kinds of interpolating currents to investigate the process. This transition had been studied in the literature by several authors, employing flavor symmetry or
quark model or both in Refs. \cite{Gavela,Perez,Singleton,Cheng,Migura}. There are also QCD sum rule description of the form factors \cite{Dosch}, upon which the
total decay rate are obtained.

The paper is organized as follows: In Sec. \ref{sec2}, we present
the relevant $\Lambda$ baryon DAs. Following that Sec. \ref{sec3} is
devoted to the LCSRs for the semi-leptonic
$\Lambda_c\rightarrow\Lambda\ell^+\nu$ decay form factors with two
kinds of interpolating currents for the $\Lambda_c$ baryon. The
numerical analysis and our conclusion are presented in Sec.
\ref{sec4}.
\section{The $\Lambda$ baryon Distribution Amplitudes}\label{sec2}

The DAs presented in this subsection is the same as that in our
previous work \cite{Wang} and a part of the complete results in Ref.
\cite{DAs}. Herein we only give them out for the completeness of the
paper. Our discussion for the $\Lambda$ baryon DAs parallels with
that for the nucleon in Ref. \cite{hitwist}, so we just list the
results following from that procedure and it is recommended to
consult the original paper for details. The $\Lambda$ baryon DAs are
defined through the following matrix element:
\begin{eqnarray}
&&4\langle0|\epsilon_{ijk}u_\alpha^i(a_1 x)d_\beta^j(a_2 x)s_\gamma^k(a_3 x)|P\rangle=\mathcal{A}_1(\rlap/P\gamma_5 C)_{\alpha\beta}\Lambda_{\gamma}+
\mathcal{A}_2M(\rlap/P\gamma_5C)_{\alpha\beta}(\rlap/x\Lambda)_{\gamma}  + \mathcal{A}_3M(\gamma_\mu\gamma_5 C)_{\alpha\beta}(\gamma^\mu
\Lambda)_{\gamma}\nonumber\\&&{}+ \mathcal{A}_4M^2(\rlap/x\gamma_5C)_{\alpha\beta}\Lambda_{\gamma}+ \mathcal{A}_5M^2(\gamma_\mu\gamma_5
C)_{\alpha\beta}(i\sigma^{\mu\nu}x_\nu \Lambda)_{\gamma}+ \mathcal{A}_6M^3(\rlap/x\gamma_5C)_{\alpha\beta}(\rlap/x \Lambda)_{\gamma},\label{das}
\end{eqnarray}
where $\Lambda_\gamma$ designates the spinor for the $\Lambda$ baryon with momentum $P$. Since only axial-vector DAs contribute to the final sum rules of the
process, we merely present this kind of structures for simplicity. The twist classification of those calligraphic DAs is indefinite, but they can be expressed by
the ones with definite twist as
\begin{eqnarray}
&&\mathcal{A}_1=A_1, \hspace{2.4cm}2P\cdot
x\mathcal{A}_2=-A_1+A_2-A_3, \nonumber\\&& 2\mathcal{A}_3=A_3,
\hspace{2.2cm}4P\cdot x\mathcal{A}_4=-2A_1-A_3-A_4+2A_5,
\nonumber\\&& 4P\cdot x\mathcal{A}_5=A_3-A_4,\hspace{0.5cm}
(2P\cdot x)^2\mathcal{A}_6=A_1-A_2+A_3+A_4-A_5+A_6.
\end{eqnarray}
The twist classification of $A_i$ is given in Table. \ref{twist}.
\begin{table}[htb]
\begin{tabular}{|c|c|c|c|}\hline
 Twist-3 &{Twist-4}&Twist-5&Twist-6\\\hline
$A_1$&$A_2$,$A_3$ &$A_4$,$A_5$&$A_6$ \\\hline
\end{tabular}
\caption{The twist for $A_i$.} \label{twist}
\end{table}
Each distribution amplitudes $F=A_i$ can be represented as Fourier integral over the longitudinal momentum fractions $x_1$, $x_2$, $x_3$ carried by the quarks
inside the baryon with $\Sigma_i x_i=1$,
\begin{equation}
F(a_iP\cdot x)=\int \mathcal{D}x e^{-ip\cdot x\Sigma_ix_ia_i}F(x_i)\;.\nonumber
\end{equation}
The integration measure is defined as
\begin{equation}
\int\mathcal{D}x=\int_0^1dx_1dx_2dx_3\delta(x_1+x_2+x_3-1).\nonumber
\end{equation}
As elucidated in Ref. \cite{Braun99}, those distribution amplitudes are scale dependent and can be expanded into orthogonal functions with increasing conformal
spin. To the leading conformal spin, or s-wave, accuracy the explicit expansions read \cite{hitwist,DAs}
\begin{eqnarray}
A_1(x_i,\mu)&=&-\,120x_1x_2x_3\phi_3^0(\mu),\nonumber\\
A_2(x_i,\mu)&=&-\,24x_1x_2\phi_4^0(\mu),\nonumber\\
A_3(x_i,\mu)&=&-\,12x_3(1-x_3)\psi_4^0(\mu)
,\nonumber\\
A_4(x_i,\mu)&=&-\,3(1-x_3)\psi_5^0(\mu),\nonumber\\
A_5(x_i,\mu)&=&-\,6x_3\phi_5^0(\mu)\nonumber\\
A_6(x_i,\mu)&=&-\,2\phi_6^0(\mu), \label{da-a}
\end{eqnarray}
where the constraint $A(x_1,x_2,x_3)=A(x_2,x_1,x_3)$ has been used in the derivation, which arises from the fact that the $\Lambda$ baryon has isospin $0$. All the
six parameters involved in Eq. (\ref{da-a}) can be expressed in terms of two independent matrix elements of local operators. Those parameters are expressed as
\begin{eqnarray}
\phi_3^0 = \phi_6^0 = -f_\Lambda,  \qquad \phi_4^0 = \phi_5^0 =
-\frac{1}{2} \left(\lambda_1 + f_\Lambda\right), \qquad
\psi_4^0 = \psi_5^0 = -\frac{1}{2}\left(\lambda_1-f_\Lambda
\right) \,. \nonumber
\end{eqnarray}
The nonperturbative parameter $f_\Lambda$ originates from the
following local matrix element:
\begin{equation}\label{fl}
\langle 0\mid \epsilon_{ijk}[u^i(0)C\gamma_5\rlap/ zd^j(0)]\rlap/
z s^k(0)\mid P\rangle=f_\Lambda z\cdot P\rlap/ z \Lambda(P).
\end{equation}
The remaining parameter $\lambda_1$ is defined by the matrix
element
\begin{equation}\label{l1}
\langle 0\mid \epsilon_{ijk}[u^i(0)C\gamma_5\gamma_\mu
d^j(0)]\gamma^\mu s^k(0)\mid P\rangle=\lambda_1 M \Lambda(P).
\end{equation}

\section{$\Lambda_c\rightarrow\Lambda\ell^+\nu$ decay form factors from light-cone Sum Rules}
\label{sec3}

\subsection{LCSR with CZ-type current}

In compliance with the standard philosophy of the LCSR method, we start from the analysis of the following correlation function
\begin{equation}
\label{correlator}
 z^\nu T_\nu(P,q)=iz^\nu\int d^4xe^{iq\cdot x}\langle 0\mid T\{j_{\Lambda_c}(0) j_\nu(x)\} \mid P\rangle,
\end{equation}
where $j_{\Lambda_c}=\epsilon_{ijk}(u^i C\gamma_5\rlap/z d^j)\rlap/z
c^k$ is the current interpolating the $\Lambda_c$ baryon state which
is similar to that used for $J^P=\frac12^+$ octet baryons by
Chernyak, Ogloblin and Zhitnitsky (CZ-type current) \cite{Chernyak},
$j_ \nu=\bar c\gamma_\nu(1-\gamma_5)s$ is the weak current, $C$ is
the charge conjugation matrix, and $i$, $j$, $k$ denote the color
indices. The auxiliary light-cone vector $z$ is introduced to
project out the main contribution onto the light cone. The coupling
constant of the baryonic current to the vacuum can thus be defined
as
\begin{equation}
\langle 0\mid j_{\Lambda_c} \mid
\Lambda_c(P')\rangle=f_{\Lambda_c}z\cdot P'\rlap/z\,\Lambda_c(P'),
\label{flc}
\end{equation}
where $\Lambda_c(P')$ and $P'$ is the $\Lambda_c$ baryon spinor and four-momentum, respectively. Form factors are defined in the usual way
\begin{eqnarray}
\langle\Lambda_c(P-q)\mid j_\nu \mid \Lambda(P) \rangle&=&\bar \Lambda_c(P-q)\left[f_1\gamma_\nu-i\frac{f_2}{M_{\Lambda_c}}\sigma_{\nu\mu}q^\mu
\right.\nonumber\\&-&\left.\left(g_1\gamma_\nu+i\frac{g_2}{M_{\Lambda_c}} \sigma_{\nu\mu}q^\mu\right)\gamma_5\right]\Lambda(P),\label{ff}
\end{eqnarray}
in which $M_{\Lambda_c}$ is the $\Lambda_c$ mass, $\Lambda(P)$ denotes the $\Lambda$ spinor and satisfies $\rlap/P\Lambda(P)=M\Lambda(P)$ with $M$ the $\Lambda$
mass and $P$ its four-momentum. The form factors that give no contribution in the case of massless final leptons are omitted here.

With those definitions (\ref{flc}) and (\ref{ff}), the hadronic representation of the correlation function (\ref{correlator}) can be written as
\begin{equation}\label{ztha}
z^\nu T_\nu=\frac{2f_{\Lambda_c}}{M_{\Lambda_c}^2-P'^2}(z\cdot
P')^2\left[f_1\rlap/z+f_2\frac{\rlap/z\rlap/q}{M_{\Lambda_c}}-
\left(g_1\rlap/z-g_2\frac{\rlap/z\rlap/q}{M_{\Lambda_c}}\right)\gamma_5\right]\Lambda(P)+\cdots,
\end{equation}
where $P'=P-q$ and the dots stand for the higher resonances and continuum contributions. While on the theoretical side, at large Euclidean momenta $P'^2$ and $q^2$
the correlation function (\ref{correlator}) can be calculated perturbatively to the leading order of the QCD coupling $\alpha_s$:
\begin{equation}\label{ztth}
z^\nu T_\nu=-2(C\gamma_5\rlap/z)_{\alpha\beta}(\rlap/z(1-\gamma_5))_\mu\int d^4x\int\frac{d^4k}{(2\pi)^4}\frac{z\cdot k}{k^2-m_c^2}\;e^{i(k+q)\cdot x}\;\langle
0\mid \epsilon_{ijk}u^i_\alpha(0)d^j_\beta(0) s^k_\mu(x)\mid P\rangle\;,
\end{equation}
where $m_c$ is the $c$-quark mass. Substituting (\ref{das}) into Eq. (\ref{ztth}) we obtain
\begin{eqnarray}
z^\nu T_\nu&=&-2(z\cdot P)^2\left[\int dx_3\;
\frac{x_3B_0(x_3)}{k^2-m_c^2}
+M^2\int dx_3\;\frac{x_3^2B_1(x_3)}{(k^2-m_c^2)^2}+2M^4\int
dx_3\;\frac{x_3^3B_2(x_3)}{(k^2-m_c^2)^3}\right]
\rlap/z(1-\gamma_5)\Lambda(P) \nonumber\\&&+2(z\cdot
P)^2\left[M\int dx_3\;\frac{x_3B_3(x_3)}{(k^2-m_c^2)^2}+2M^3\int
dx_3\;\frac{x_3^2B_2(x_3)}{(k^2-m_c^2)^3}\right] \rlap/z\rlap/q
(1+\gamma_5)\Lambda(P)+\cdots,\label{ztres}
\end{eqnarray}
where $k=x_3P-q$ and the ellipses stand for contributions that are
nonleading in the infinite momentum frame kinematics
$P\rightarrow\infty$, $q\sim \mbox{const.}$, $z\sim 1/P$. The
functions $B_i$ are defined by
\begin{eqnarray}
B_0(\alpha_3)&=&\int_0^{1-x_3}dx_1A_1(x_1,1-x_1-x_3,x_3),\nonumber\\
B_1(\alpha_3)&=&-2\tilde{A_1}+\tilde{A_2}-\tilde{A_3}-\tilde{A_4}
+\tilde{A_5},\nonumber\\
B_2(\alpha_3)&=&\tilde{\tilde{A_1}}-\tilde{\tilde{A_2}} +\tilde{\tilde{A_3}}+\tilde{\tilde{A_4}}-\tilde{\tilde{A_5}}
+\tilde{\tilde{A_6}},\nonumber\\
B_3(\alpha_3)&=&-\tilde{A_1}+\tilde{A_2}-\tilde{A_3}.
\end{eqnarray}
The DAs with tildes are defined via integration as follows
\begin{eqnarray}
\tilde{A}(x_3)&=&\int_1^{x_3}dx'_3\int_0^{1-x'_3}dx_1A(x_1,1-x_1-x'_3,x'_3),\nonumber\\
\tilde{\tilde{A}}(x_3)&=&\int_1^{x_3}dx'_3\int_1^{x'_3}dx''_3\int_0^{1-x''_3}dx_1
A(x_1,1-x_1-x''_3,x''_3).
\end{eqnarray}
These functions originate from the partial integration, which is
used to eliminate the factors $1/(P\cdot x)^n$ appearing in the
distribution amplitudes. When the next-to-leading order conformal
expansion is considered, the surface terms completely sum to zero.
The term $B_0$ corresponds to the leading twist contribution. The
form factors $f_2$ and $g_2$ in Eq. (\ref{ztres}) are characterized
by the higher twist contributions.

Matching Eqs. (\ref{ztha}) and (\ref{ztres}), adopting the quark-hadron duality assumption and employing a Borel improvement of $P'^2$ on both sides lead us to the
desired sum rules for the form factors $f_1$ and $f_2$:
\begin{subequations}
\label{sr-CZ}
\begin{eqnarray}\label{sr-f1}
-f_{\Lambda_c} f_1 e^{-M_{\Lambda_c}^2/M_B^2}&=&-\int_{x_0}^1dx_2\;e^{-s'/M_B^2}\left[ B_0+\frac{M^2}{M_B^2}\left(-B_1(x_3)
+\frac{M^2}{M_B^2}B_2(x_3)\right)\right]\nonumber\\
&&+\frac{M^2x_0^2e^{-s_0/M_B^2}}{m_c^2-q^2+x_0^2 M^2}\left(B_1(x_0) -\frac{M^2}{M_B^2}B_2(x_0)\right)\nonumber\\
&&+\frac{M^2e^{-s_0/M_B^2}x_0^2} {m_c^2-q^2+x_0^2M^2} \frac{d}{dx_0}\left(\frac{M^2 x_0^2B_2(x_0) }{m_c^2-q^2+x_0^2M^2} \right) ,
\\
\nonumber\\
\label{sr-f2} \frac{f_{\Lambda_c} f_2}{M_{\Lambda_c}M} e^{-M_{\Lambda_c}^2/M_B^2}&=&\frac{1}{M_B^2}\int_{x_0}^1\frac{dx_3}{x_3}\;e^{-s'/M_B^2}\left(
B_3(x_3)-\frac{M^2}{M_B^2}B_2(x_3)\right)\nonumber\\
&&+\frac{x_0e^{-s_0/M_B^2}}{m_c^2-q^2+x_0^2M^2} \left(B_3(x_0)-\frac{M^2}{M_B^2}B_2(x_0)\right)\nonumber\\
&&+\frac{M^2e^{-s_0/M_B^2}x_0^2}{m_c^2-q^2+x_0^2M^2} \frac{d}{dx_0}\left(\frac{x_0B_2(x_0)}{m_c^2-q^2+x_0^2M^2}\right),
\end{eqnarray}
\end{subequations}
where
\begin{equation}
s'=(1-x)M^2+\frac{m_c^2-(1-x)q^2}{x},\nonumber
\end{equation}
and $x_0$ is the positive solution of the quadratic equation for
$s'=s_0$:
\begin{equation}
2M^2x_0=\sqrt{(-q^2+s_0-M^2)^2+4M^2(-q^2+m_c^2)}-(-q^2+s_0-M^2).\nonumber
\end{equation}
As the sum rules for the form factors $g_1$ and $g_2$ are identical with those for the $f_1$ and $f_2$, $f_1=g_1$ and $f_2=g_2$, we will only discuss the results
for $f_1$ and $f_2$ in the numerical analysis section.
\subsection{LCSR with Ioffe-type interpolating current}
The interpolating current used for the baryon is not the unique one.
As exemplified in the studies \cite{current} for the applications of
QCD sum rules, there are other choices existing in interpolating
baryonic state with the same quantum numbers. It has been known that
the current interpolating the hadron state is of great importance in
field theory and can affect the calculations to some extent
\cite{EMff,Lee,IIG}. In this subsection, we adopt another
interpolating current to investigate the form factors of the
semi-leptonic decay $\Lambda_c\rightarrow\Lambda\ell^+\nu$. To this
end, the Ioffe-type baryonic current (see \cite{current} for a
reference) is used in the calculation, that is
$j_{\Lambda_c}=\epsilon_{ijk}(u^i C\gamma_5\gamma_\mu d^j)\gamma^\mu
c^k$ . The coupling of this kind of interpolating current is
determined by the following matrix element:
\begin{equation}
\langle 0\mid j_{\Lambda_c} \mid \Lambda_c(P')\rangle=\lambda_{1c}M_{\Lambda_c}\Lambda_c(P'). \label{lamlc}
\end{equation}

In order to derive the LCSR of the weak transition form factors, we need to express the correlation function (\ref{correlator}) both phenomenologically and
theoretically. By inserting a complete set of intermediate states with the same quantum numbers as those of $\Lambda_c$, the hadronic representation of the
correlation function is expressed as
\begin{eqnarray}
z^\nu T_\nu&=&\frac{\lambda_{1c}M_{\Lambda_c}}{M_{\Lambda_c}^2-P'^2}\Big [2P\cdot zf_1(q^2)+\frac{2P\cdot z}{M_{\Lambda_c}}\rlap/q_\perp f_2(q^2)-[M_\Lambda
f_1(q^2)-\frac{q^2}{M_{\Lambda_c}}f_2(q^2)-M_{\Lambda_c}f_1(q^2)]\rlap/z\nonumber\\&& +[f_1(q^2)+f_2(q^2)+\frac{M_\Lambda}{M_{\Lambda_c}}f_2(q^2)]\rlap/z\rlap/q
-2P\cdot zg_1(q^2)\gamma_5+\frac{2P\cdot z}{M_{\Lambda_c}}g_2(q^2)\rlap/q_\perp\gamma_5\nonumber\\
&&-[M_\Lambda
g_1(q^2)-\frac{q^2}{M_{\Lambda_c}}g_2(q^2)+M_{\Lambda_c}g_1(q^2)]\rlap/z\gamma_5-[g_1(q^2)-g_2(q^2)+\frac{M_\Lambda}{M_{\Lambda_c}}g_2(q^2)]\rlap/z\rlap/q\gamma_5]\Lambda(P)
+...,
\end{eqnarray}
where $\rlap/q_\perp=\rlap/q-\frac{p\cdot q}{p\cdot z}\rlap/z$, and
the dots stand for the higher resonance and continuum contributions.
In contrast to the case of CZ-type current, there are more Lorentz
structures existing in the expression. Generally speaking, all of
the Lorenz structures can give information on the form factors. In
this paper we choose the terms proportional to $1$ and
$\rlap/q_\perp$ to get the sum rules. After the standard procedure
of LCSR method, we arrive at the following sum rules:
\begin{subequations}
\label{sr-ioffe}
\begin{eqnarray}
\label{srf1-ioffe}
\lambda_{1c}M_{\Lambda_c}f_1(q^2)&=&\int_{\alpha_{30}}^1d\alpha_3e^{-\frac{s-M_{\Lambda_c}^2}{M_B^2}}\Big\{-\frac{m_c}{\alpha_3}B_0(\alpha_3)+MB_0'(\alpha_3)
+\frac{M}{\alpha_3M_B^2}(M_B^2+\frac{q^2}{\alpha_3}+Mm_c)B_3(\alpha_3)\nonumber\\
&&+\frac{M^2m_c}{\alpha_3M_B^2}B_5(\alpha_3)+\frac{M^2}{\alpha_3M_B^2}(\alpha_3M+m_c)B_4(\alpha_3)
-\frac{M^4m_c}{\alpha_3M_B^4}B_2(\alpha_3)\Big\}\nonumber\\
&&+e^{-\frac{s_0-M_{\Lambda_c}^2}{M_B^2}}\frac{1}{\alpha_{30}^2M^2+m_c^2-q^2}\Big\{(Mq^2+\alpha_{30}M^2m_c)B_3(\alpha_{30})+\alpha_{30}M^2m_cB_5(\alpha_{30})\nonumber\\
&&+M^2(\alpha_{30}^2M+\alpha_{30}m_c)B_4(\alpha_{30})-\frac{\alpha_{30}M^4m_c}{M_B^2}B_2(\alpha_{30})\Big\}\nonumber\\
&&+e^{-\frac{s_0-M_{\Lambda_c}^2}{M_B^2}}\frac{\alpha_{30}^2}{\alpha_{30}^2M^2+m_c^2-q^2}\frac{d}{d\alpha_{30}}\frac{\alpha_{30}M^4m_c}{\alpha_{30}^2M^2+m_c^2-q^2}B_2(\alpha_{30}),
\\
\label{srf2-ioffe}
\lambda_{1c}f_2(q^2)&=&\int_{\alpha_{30}}^1d\alpha_3e^{-\frac{s-M_{\Lambda_c}^2}{M_B^2}}\Big\{\frac{1}{\alpha_3}B_0(\alpha_3)-\frac{M}{\alpha_3M_B^2}(M+\frac{m_c}{\alpha_3})B_3(\alpha_3)
-\frac{M^2}{\alpha_3M_B^2}B_4(\alpha_3)+\frac{M^3m_c}{\alpha_3^2M_B^4}B_2(\alpha_3)\Big\}\nonumber\\
&&-e^{-\frac{s_0-M_{\Lambda_c}^2}{M_B^2}}\frac{1}{\alpha_{30}^2M^2+m_c^2-q^2}\Big\{(\alpha_{30}M^2+Mm_c)B_3(\alpha_{30})+\alpha_{30}M^2B_4(\alpha_{30})
-\frac{M^3m_c}{M_B^2}B_2(\alpha_{30})\Big\}\nonumber\\
&&-e^{-\frac{s_0-M_{\Lambda_c}^2}{M_B^2}}\frac{\alpha_{30}^2}{\alpha_{30}^2M^2+m_c^2-q^2}\frac{d}{d\alpha_{30}}\frac{M^3m_c}{\alpha_{30}^2M^2+m_c^2-q^2}B_2(\alpha_{30}),
\\
\label{srg1-ioffe}
\lambda_{1c}M_{\Lambda_c}g_1(q^2)&=&\int_{\alpha_{30}}^1d\alpha_3e^{-\frac{s-M_{\Lambda_c}^2}{M_B^2}}\Big\{-\frac{m_c}{\alpha_3}B_0(\alpha_3)-MB_0'(\alpha_3)
-\frac{M}{\alpha_3M_B^2}(\frac{q^2}{\alpha_3}+M_B^2-Mm_c)B_3(\alpha_3)\nonumber\\
&&+\frac{M^2m_c}{\alpha_3M_B^2}B_5(\alpha_3)+\frac{M^2}{\alpha_3M_B^2}(-\alpha_3M+m_c)B_4(\alpha_3)
-\frac{M^4m_c}{\alpha_3M_B^4}B_2(\alpha_3)\Big\}\nonumber\\
&&-e^{-\frac{s_0-M_{\Lambda_c}^2}{M_B^2}}\frac{1}{\alpha_{30}^2M^2+m_c^2-q^2}\Big\{(Mq^2-\alpha_{30}M^2m_c)B_3(\alpha_{30})-\alpha_{30}M^2m_cB_5(\alpha_{30})\nonumber\\
&&+M^2(\alpha_{30}^2M-\alpha_{30}m_c)B_4(\alpha_{30})+\frac{\alpha_{30}M^4m_c}{M_B^2}B_2(\alpha_{30})\Big\}\nonumber\\
&&+e^{-\frac{s_0-M_{\Lambda_c}^2}{M_B^2}}\frac{\alpha_{30}^2}{\alpha_{30}^2M^2+m_c^2-q^2}\frac{d}{d\alpha_{30}}\frac{\alpha_{30}M^4m_c}{\alpha_{30}^2M^2+m_c^2-q^2}B_2(\alpha_{30}),
\\
\label{srg2-ioffe}
\lambda_{1c}g_2(q^2)&=&\int_{\alpha_{30}}^1d\alpha_3e^{-\frac{s-M_{\Lambda_c}^2}{M_B^2}}\Big\{-\frac{1}{\alpha_3}B_0(\alpha_3)+\frac{M}{\alpha_3M_B^2}(M-\frac{m_c}{\alpha_3})B_3(\alpha_3)
+\frac{M^2}{\alpha_3M_B^2}B_4(\alpha_3)+\frac{M^3m_c}{\alpha_3^2M_B^4}B_2(\alpha_3)\Big\}\nonumber\\
&&+e^{-\frac{s_0-M_{\Lambda_c}^2}{M_B^2}}\frac{1}{\alpha_{30}^2M^2+m_c^2-q^2}\Big\{(\alpha_{30}M^2-Mm_c)B_3(\alpha_{30})+\alpha_{30}M^2B_4(\alpha_{30})
+\frac{M^3m_c}{M_B^2}B_2(\alpha_{30})\Big\}\nonumber\\
&&-e^{-\frac{s_0-M_{\Lambda_c}^2}{M_B^2}}\frac{\alpha_{30}^2}{\alpha_{30}^2M^2+m_c^2-q^2}\frac{d}{d\alpha_{30}}\frac{M^3m_c}{\alpha_{30}^2M^2+m_c^2-q^2}B_2(\alpha_{30}),
\end{eqnarray}
\end{subequations}
where the additional definitions are used as
\begin{eqnarray}
B_0'(\alpha_3)&=&\int_0^{1-x_3}dx_1A_3(x_1,1-x_1-x_3,x_3),\nonumber\\
B_4(\alpha_3)&=&\tilde{A_3}-\tilde{A_4},\nonumber\\
B_5(\alpha_3)&=&-\tilde{A_1}-\tilde{A_3}+\tilde{A_5}.
\end{eqnarray}
\section{numerical analysis and the conclusion}\label{sec4}

\subsection{Values for $f_\Lambda$ and $\lambda_1$}

The explicit expressions of the baryon DAs rely on the
nonperturbative parameters $f_\Lambda$ and $\lambda_1$, which need
to be determined by some nonperturbative method. In the present
subsection we use the QCD sum rule approach to estimate the values.
According to their definitions, we consider correlation functions
\begin{equation}
\label{cf-corr} \Pi(q^2)=i\int d^4xe^{iq\cdot x}\langle 0\mid T\{J_i(x) \bar J_j(0)\} \mid 0\rangle,
\end{equation}
where $J_i$ are currents given in (\ref{fl}) and (\ref{l1}). Following the standard procedure, the sum rules of $f_\Lambda$,$\lambda_1$ and their relative sign are
straightforward
\begin{eqnarray}\label{f1-sr}
(4\pi)^4f_\Lambda^2e^{-M^2/M_B^2}&=&\frac{2}{5}\int_{{m_s^2}}^{s_0}
s(1-x)^5e^{-s/M_B^2}ds-\frac{b}{3}\int_{{m_s^2}}^{s_0}
x(1-x)(1-2x)e^{-s/M_B^2}\frac{ds}{s},
\end{eqnarray}
\begin{eqnarray}\label{f2-sr}
4(2\pi)^4\lambda_1^2M^2e^{-M^2/M_B^2}&=&\frac12\int_{m_s^2}^{s_0}s^2[(1-x)(1+x)(1-8x+x^2)-12x^2\ln
x]e^{-s/M_B^2}ds\nonumber\\
&&+\frac{b}{12}\int_{m_s^2}^{s_0}(1-x)^2e^{-s/M_B^2}ds-\frac43a^2(1-\frac{m_0^2}{2M_B^2}-\frac{m_0^2m_s^2}{2M_B^4}\nonumber\\
&& +\frac{m_0^4m_s^4}{16M_B^8})e^{-m_s^2/M_B^2}-m_sa_s\int_{m_s^2}^{s_0}e^{-s/M_B^2}ds,
\end{eqnarray}
\begin{eqnarray}\label{sign-sr}
(4\pi)^4f_\Lambda\lambda_1^*Me^{-M^2/M_B^2}=\frac23m_s\int_{m_s^2}^{s_0}s[(1-x)(3+13x-5x^2+x^3)+12x\ln
x]e^{-s/M_B^2}ds\nonumber\\
+\frac b3m_s\int_{m_s^2}^{s_0}\frac1s(1-x)[1+\frac{(1-x)(2-5x)}{3x}]e^{-s/M_B^2}ds+\frac{8}{3}a_s\int_{m_s^2}^{s_0}e^{-s/M_B^2}ds.
\end{eqnarray}
where $x=m_s^2/s$ and $m_s$ is the $s$-quark mass. The sum rule for the $f_\Lambda$ has been obtained before \cite{lbtop} where the corresponding heavy quark limit
is also derived. At the working window $s_0\sim2.55\;\mbox{GeV}^2$ and $1<M_B^2<2\;\mbox{GeV}^2$ the numerical values for the coupling constants read
\begin{eqnarray}
f_\Lambda=6.0\times10^{-3}\mbox{GeV}^2,\hspace{0.3cm} \lambda_1=1.0
\times10^{-2} \mbox{GeV}^2.
\end{eqnarray}
The relative sign of $\lambda_1$ and $f_\Lambda$ is obtained from the sum rule (\ref{sign-sr}). In the numeric analysis, the standard values $a=-(2\pi)^2\langle
\bar qq \rangle=0.55\mbox{GeV}^3$, $a_s=-(2\pi)^2\langle \bar ss \rangle=0.8a$, $b=(2\pi)^2\langle \alpha_sG^2/\pi \rangle=0.47~\mbox{GeV}^4$ and
$m_s=0.15\mbox{GeV}$ are adopted. It should be noted that our value for $f_\Lambda$ here does coincide with that obtained in Ref. \cite{Chernyak}. In comparison
with the previous work \cite{Wang}, sum rules of $\lambda_1^2$ and $f_\Lambda \lambda_1$ are renewed with the consideration of four-quark condensate contributions
$a^2$. The numerical analysis shows that the renewed sum rules give a different sign to the parameter $\lambda_1$, which may lead to the correct results on the
electromagnetic form factors \cite{EMff}.

The sum rules for the coupling of $\Lambda_c$ to vacuum,
$f_{\Lambda_c}$ and $\lambda_{1c}$, are similar to that for
$\Lambda$, where the simple substitution $m_s\rightarrow m_c$ should
be made. The numerical estimates are
$f_{\Lambda_c}=(9.1\pm0.5)\times 10^{-3}\mbox{GeV}^2$ and
$\lambda_{1c}=(1.2\pm0.2)\times 10^{-2}\mbox{GeV}^2$, taken from the
interval $2<M_B^2<3\mbox{GeV}^2$ for $f_{\Lambda_c}$ with
$s_0\sim10\,\mbox{GeV}^2$ and $1.5<M_B^2<2.0\,\mbox{GeV}^2$ for
$\lambda_{1c}$ with $s_0\sim9\,\mbox{GeV}^2$.

\subsection{Analysis of the LCSRs}
\begin{figure}[t]
\centerline{\epsfysize=6truecm \epsfbox{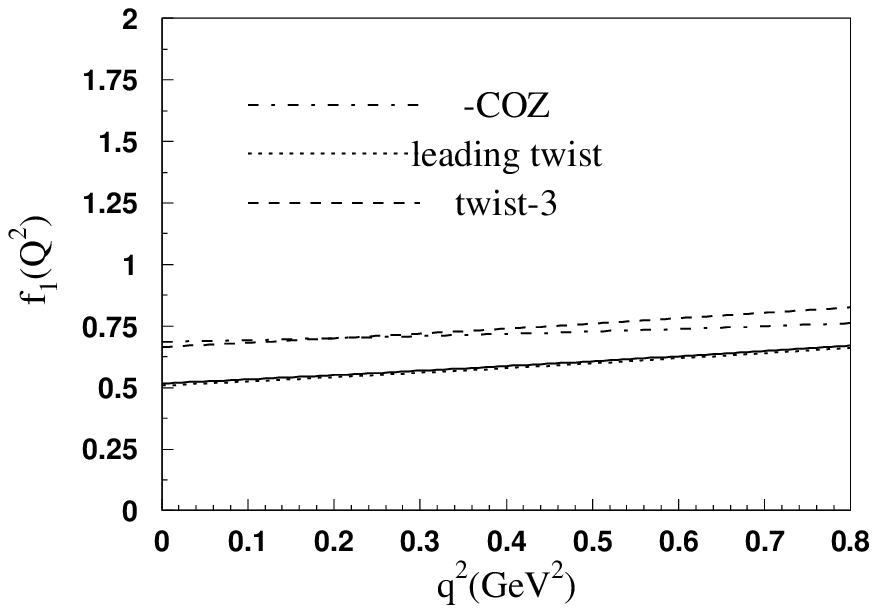}
  \epsfysize=6truecm \epsfbox{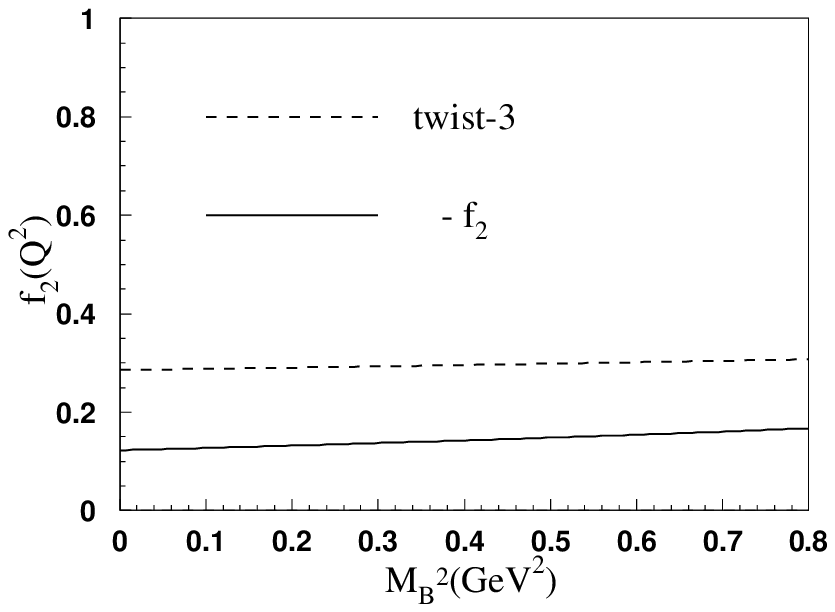}}
\caption{The dependence on $q^2$ of the LCSRs for the form factors $f_1$ and $f_2$. The ``COZ" denotes the result obtained from the COZ DAs. The continuum threshold
and the Borel parameter are $s_0=10\mbox{GeV}^2$ and $M_B^2=9\mbox{GeV}^2$.} \label{fig:2}
\end{figure}

In the numerical analysis for the sum rules of the form factors, the
charm quark mass is taken to be $m_c=1.27~\mbox{GeV}$, and the other
relevant parameters, $\Lambda_c$ and $\Lambda$ baryon masses and the
value of $|V_{cs}|$, are used as the centra values provided by PDG
\cite{PDG}. The analysis starts from the sum rules with the CZ-type
interpolating current. We first take into account contributions from
the twist-3 DAs, in which only twist-3 DA is kept. Substituting the
above given parameters into the LCSRs and varying the continuum
threshold within the range $s_0=9-11~\mbox{GeV}^2$, we find there
exist an acceptable stability in the working window
$M_B^2=8-10~\mbox{GeV}^2$ for the Borel parameter. The $M_B^2$ and
the $q^2$ dependence for the
\begin{figure}[b]
\centerline{\epsfysize=6truecm \epsfbox{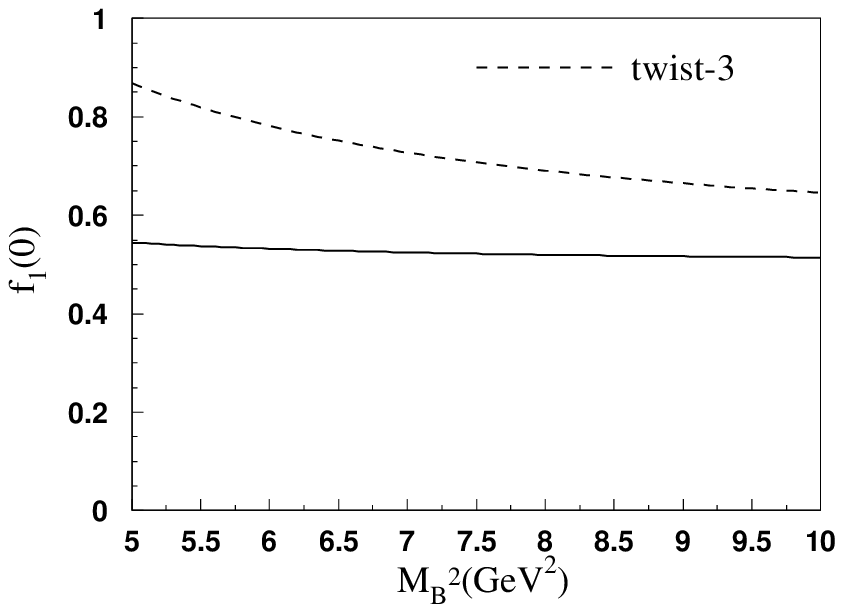} \epsfysize=6truecm \epsfbox{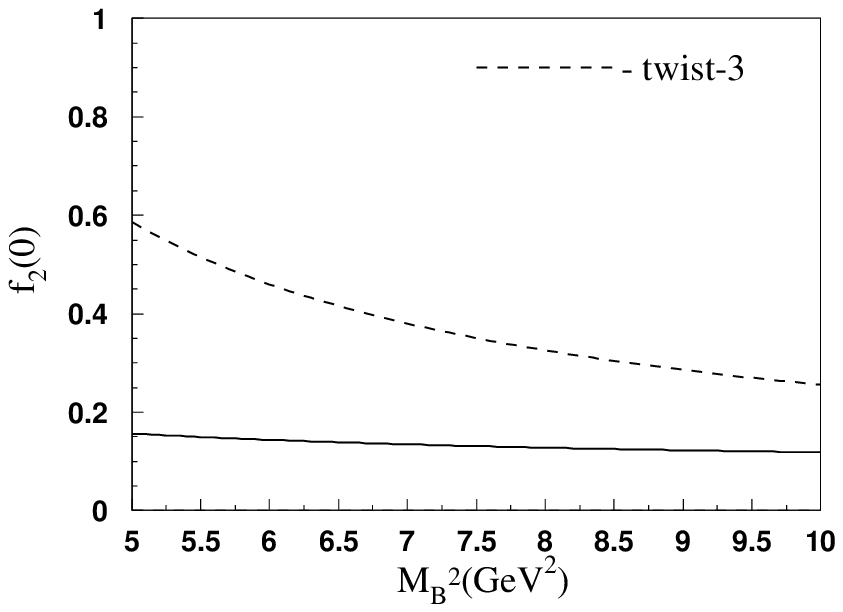}} \caption{The dependence on $M_B^2$ of the LCSRs for the form factors $f_1$
and $f_2$ at $q^2=0$. The continuum threshold is $s_0=10\mbox{GeV}^2$.} \label{fig:1}
\end{figure}
corresponding form factors are shown in Figs. \ref{fig:1} and \ref{fig:2},
respectively. Also given in Fig. \ref{fig:2} are two leading twist results, corresponding to contributions only retaining $B_0$ in the sum rules. It is apparent
that only $f_1$ survives in the approximation. Apart from the leading twist DA we discuss in Eq. (\ref{da-a}), there still exists another form from Chernyak,
Ogloblin and Zhitnitsky \cite{Chernyak}:
\begin{equation}
A^{COZ}_1(x_i)=-21\varphi_{as}[0.52(x_1^2+x_2^2)+0.68x_3^2-1.12x_1x_2-0.48x_3(x_1+x_2)], \label{coz}
\end{equation}
where $\varphi_{as}=120x_1x_2x_3$ is the asymptotic DA. The corresponding result is also illustrated for contrast.

When including contributions up to twist-6, the stability is agreeable within the range $s_0=9-11~\mbox{GeV}^2$ and $M_B^2=7-9~\mbox{GeV}^2$. In that working
region, the twist-3 contribution to $f_1$ is the dominant one, while for $f_2$, the main contribution comes from the twist-4 DAs and its magnitude is approximately
$\sim 1.5$ of the twist-3 one in the whole dynamical region, but with a different sign. On account of the relatively small momentum transfer, the asymptotic
behavior of DAs may not be fulfilled and we need incorporate higher conformal spin in the expansion for them. Furthermore, QCD sum rule tends to overestimate the
higher conformal spin expansion parameters \cite{nonlocal}, and the corresponding parameter will enter in the coefficients of the higher conformal spin expansion,
which is well known as the Wandzura-Wilczek type contribution. The $M_B^2$ and the $q^2$ dependence for the corresponding form factors are also shown in Figs.
\ref{fig:1} and \ref{fig:2}.

Both form factors can be well fitted by the three-parameter dipole formula in the LCSR allowed region, $0<q^2<0.8\,\mbox{GeV}^2$:
\begin{equation}
f_i(q^2)=\frac{f_i(0)}{a_2(q^2/M_{\Lambda_c}^2)^2+a_1q^2/M_{\Lambda_c}^2+1}. \label{dipole}
\end{equation}
Below in Table \ref{di-fit} we give the central values of those coefficients with parameters $M_B^2=9~\mbox{GeV}^2$ and $s_0=10~\mbox{GeV}^2$. It is noted that in
the table, $f_1=g_1$ and $f_2=g_2$ for the CZ-type interpolating current.
\begin{table}[htb]
\begin{tabular}{|c|c|c|c|c|c|c|c|c|c|c|c|c|c|c|}
\hline &\multicolumn{6}{|c|}{CZ-type}&\multicolumn{6}{|c|}{Ioffe-type}\\
\hline
&\multicolumn{3}{|c|}{Twist-3} &\multicolumn{3}{|c|}{Twist-6}& \multicolumn{3}{|c|}{Twist-3}&\multicolumn{3}{|c|}{Twist-6}\\
\hline    &$f_i(0)$&$a_1$&$a_2$&$f_i(0)$&$a_1$&$a_2$&$f_i(0)$&$a_1$&$a_2$&$f_i(0)$&$a_1$&$a_2$\\
\hline $f_1$&$0.665$&$-1.350$&$0.460$&$0.517$&$-1.692$&$0.941$&$-0.991$&$-2.361$&$1.521$&$-0.387$&$-1.920$&$1.915$ \\
\hline $f_2$&$-0.285$&$-0.421$&$-0.243$&$0.123$&$-1.902$&$1.080$&$1.309$&$-2.066$&$-0.0001$&$0.664$&$-2.251$&$0.648$ \\
\hline $g_1$&$0.665$&$-1.350$&$0.460$&$0.517$&$-1.692$&$0.941$&$-0.339$&$-2.481$&$0.101$&$-0.702$&$-3.219$&$2.763$\\
\hline $g_2$&$-0.285$&$-0.421$&$-0.243$&$0.123$&$-1.902$&$1.080$&$-0.847$&$-3.357$&$4.020$&$-1.208$&$-3.178$&$3.114$\\
\hline
\end{tabular}
\caption{The dipole formula fit for the form factors $f_i$ and $g_i$ with two kinds of interpolating currents.} \label{di-fit}
\end{table}

\begin{figure}[b]
\centerline{\epsfysize=6truecm \epsfbox{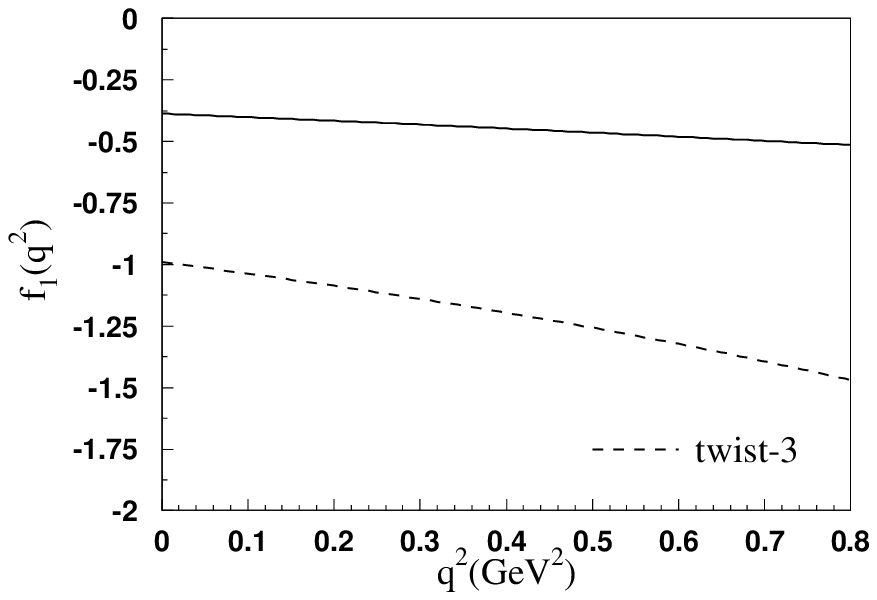} \epsfysize=6truecm \epsfbox{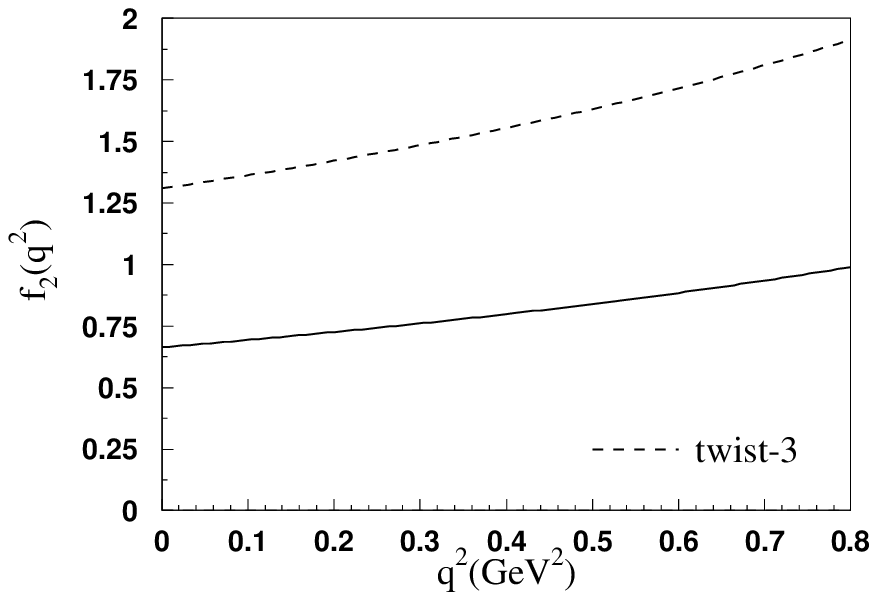}} \epsfysize=6truecm \epsfbox{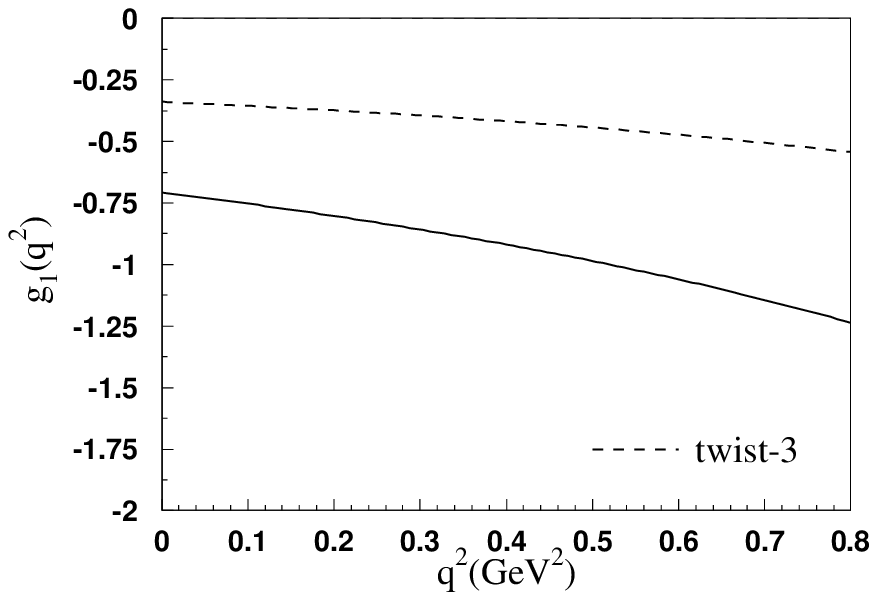} \epsfysize=6truecm
\epsfbox{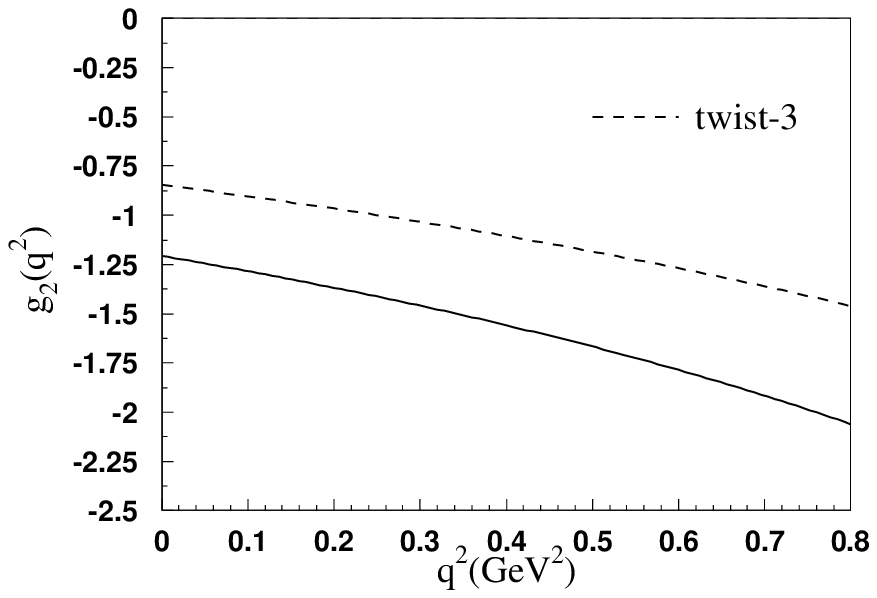} \caption{The $q^2$-dependence of the LCSRs for the form factors $f_i$ and $g_i$ with Ioffe-type interpolating current. The continuum threshold
is $s_0=9\,\mbox{GeV}^2$ and the Borel parameter is set to be $M_B^2=9\,\mbox{GeV}^2$.} \label{fig:3}
\end{figure}

For the analysis of sum rules from the Ioffe-type interpolating current, we comply with the same procedure above. The calculation shows that the form factors vary
mildly with the Borel parameter in the range $8\,\mbox{GeV}^2\le M_B^2\le 10\,\mbox{GeV}^2$ with the threshold varying in the region $8\,\mbox{GeV}^2\le s_0\le
10\,\mbox{GeV}^2$. Therefore in the following analysis we set $M_B^2=9\,\mbox{GeV}^2$. The $q^2$-dependence of the weak transition form factors are plotted in Fig.
\ref{fig:3}, with both contributions from twist-3 and up to twist-6. The figure shows that higher order twist contributions can play so important roles in the
calculation that it is necessary to include them in the investigations.
All the form factors $f_i$ and $g_i$ can be well fitted by the dipole formula (\ref{dipole}) in the range $0<q^2<0.8\,\mbox{GeV}^2$ with $s_0=9\,\mbox{GeV}^2$. The
fit coefficients are shown in Table \ref{di-fit}.

\subsection{Semileptonic decay width}

The differential decay rate of the decay process can be expressed by the weak transition form factors as
\begin{eqnarray}
\frac{d\Gamma}{dq^2}&=&\frac{G_F^2|V_{cs}|^2}{192 \pi^3M_{\Lambda_c}^5}q^2\sqrt{q_+^2q_-^2}\Big\{-6f_1f_2M_{\Lambda_c}m_{+}{q_-}^2+6g_1g_2M_{\Lambda_c}m_{-}q_+^2\nonumber\\
&&+f_1^2M_{\Lambda_c}^2(\frac{m_+^2m_-^2}{q^2}+m_-^2-2(q^2+2M_{\Lambda_c}M_\Lambda))\nonumber\\
&&+g_1^2M_{\Lambda_c}^2(\frac{m_+^2m_-^2}{q^2}+m_+^2-2(q^2-2M_{\Lambda_c}M_\Lambda))\nonumber\\
&&-f_2^2[-2m_+^2m_-^2+m_+^2q^2+q^2(q^2+4M_{\Lambda_c}M_\Lambda)]\nonumber\\
&&-g_2^2[-2m_+^2m_-^2+m_-^2q^2+q^2(q^2-4M_{\Lambda_c}M_\Lambda)]\Big\}, \label{decayrate}
\end{eqnarray}
where $m_{\pm}=M_{\Lambda_c}\pm M_\Lambda$ and $q_{\pm}=q^2-m_{\pm}^2$ are used for convenience. It has been known that the light-cone sum rules on the decay form
factors can only be reliable in the range $q^2-m_c^2\ll0$, i.e. $0<q^2<0.8\,\mbox{GeV}^2$ in our calculations, while the thorough understanding of the decay process
needs us to know information on the whole physical region $0<q^2<(M_{\Lambda_c}-M_\Lambda)^2$. In order to give numerical estimate, we extrapolate the fit formula
to the whole physical region, assuming that the form factors can be described by the dipole formula with the same coefficients in the whole kinematic range.

\begin{figure}[t]
\centerline{\epsfysize=6truecm \epsfbox{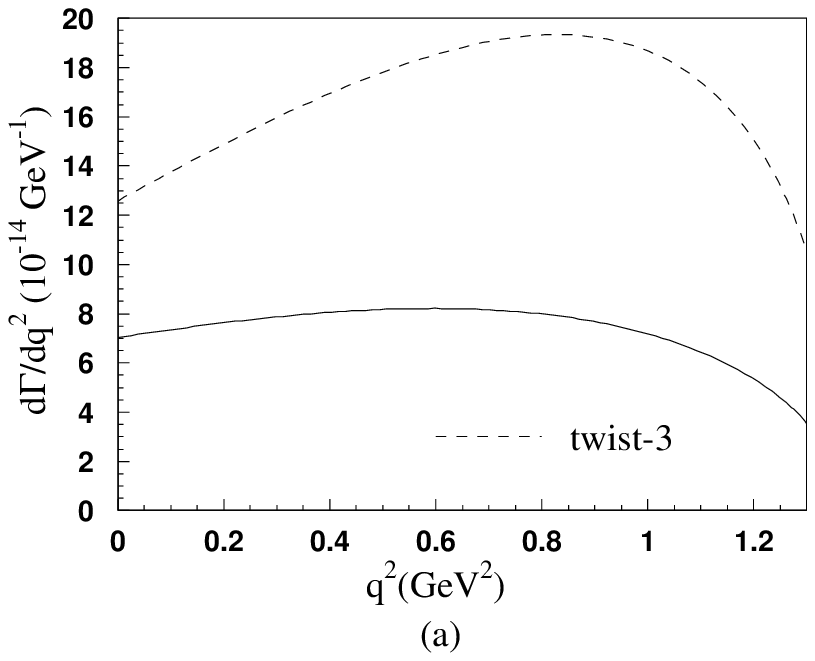} \epsfysize=6truecm \epsfbox{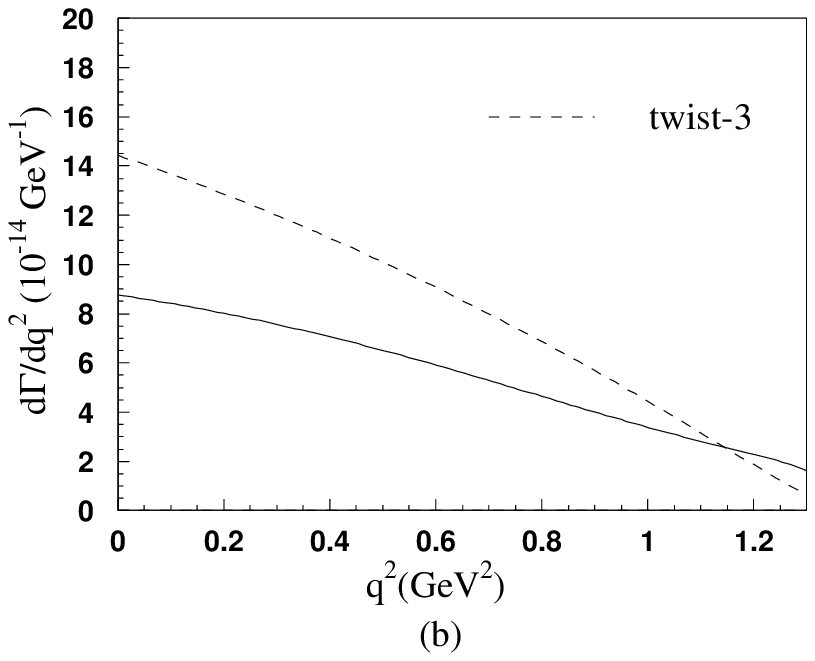}} \caption{Differential decay rate for
$\Lambda_c\rightarrow\Lambda\ell^+\nu$. (a) corresponds to the CZ-type interpolating current and (b) the Ioffe-type current.} \label{fig:4}
\end{figure}

After extrapolation of the form factors given in Table \ref{di-fit}, we calculate the differential decay rate for the process
$\Lambda_c\rightarrow\Lambda\ell^+\nu$, which is shown in Fig. \ref{fig:4} (a) for CZ-type interpolating current and Fig. \ref{fig:4} (b) for Ioffe-type current.
The estimate of the total decay width can be obtained after integration of the momentum transfer $q^2$ in the whole kinematical region. Taking into account
contributions up to twist-6 DAs, we give the prediction $\Gamma(\Lambda_c\rightarrow \Lambda\ell^+\nu)=(10.04\pm0.88)\times10^{-14}\mbox{GeV}$ for CZ-type current
and $\Gamma(\Lambda_c\rightarrow \Lambda\ell^+\nu)=(6.45\pm1.06)\times10^{-14}\mbox{GeV}$ for Ioffe-type current. The errors in our numerical results come from the
different choice of the threshold $s_0=9-11\,\mbox{GeV}^2$ with the Borel parameter varying in the region $7\,\mbox{GeV}^2\le M_B^2\le 9\,\mbox{GeV}^2$ for CZ-type
current and $s_0=8-10\,\mbox{GeV}^2$ with $8\,\mbox{GeV}^2\le M_B^2\le 10\,\mbox{GeV}^2$ for Ioffe-type current. Note that, as the estimations are from the dipole
formula fits with the sum rules as the input data, the uncertainty due to the variation of the input parameters, such as $f_{\Lambda}$ and $\lambda_1$, as well as
$f_{\Lambda_c}$ and $\lambda_{1c}$, is not included, which may reach 5-10\% or more.

For a comparison with experiments, we turn to data provided by PDG.
By use of the mean lifetime of $\Lambda_c$:
$\tau=200\times10^{-15}\,s$, the branching ratio of the process is
estimated to be $Br(\Lambda_c\rightarrow
\Lambda\ell^+\nu)=0.030\pm0.003$ for CZ-type interpolating current
and $Br(\Lambda_c\rightarrow \Lambda\ell^+\nu)=0.020\pm0.003$ for
Ioffe-type interpolating current, which are explicitly shown in
Table \ref{br}. In order to see influence of higher twist DAs to the
decay, we also give predictions from the twist-3 contributions for
the two kinds of interpolating currents in the table. The table
shows that the adoption of Ioffe-type interpolating current is a
better choice for the study of the semi-leptonic decay mode
$\Lambda_c\rightarrow\Lambda\ell^+\nu$, and the inclusion of higher
twist contributions is necessary for calculations. It can also be
seen that the improvement of the parameters in the distribution
amplitudes results in marked difference from results in \cite{Wang}
for the CZ-type current case.
\begin{table}[htb]
\begin{tabular}{|c|c|c|c|c|c|c|}
\hline &\multicolumn{3}{|c|}{$\Gamma(\times10^{-14})\mbox{GeV}$} &\multicolumn{3}{|c|}{$Br(\%)$}\\
\hline
& Twist-3 & Twist-6& PDG& Twist-3 & Twist-6& PDG\\
\hline
CZ-type&$19.7$ &$10.04$&$6.77$&$5.8$ &$3.0$&$2.0$ \\
\cline{1-1}\cline{2-2}\cline{3-3}\cline{5-5}\cline{6-6}
Ioffe-type&$10.5$&$6.5$&&$3.2$&$2.0$&\\
\hline
\end{tabular}
\caption{The prediction of the semi-leptonic decay $\Lambda_c\rightarrow \Lambda\ell^+\nu$ with two kinds of interpolating currents.} \label{br}
\end{table}

To summarize, we have given an approved investigation on the
semi-leptonic decay $\Lambda_c\rightarrow \Lambda\ell^+\nu$. The
form factors characterizing the process are studied within the
framework of the LCSR method. In the calculations we adopt both
CZ-type and Ioffe-type current as interpolating field for the
$\Lambda_c$ baryon. Light-cone sum rules for the form factors are
derived and used to predict the decay width, which is
$\Gamma(\Lambda_c\rightarrow
\Lambda\ell^+\nu)=(10.04\pm0.88)\times10^{-14}\mbox{GeV}$ from
CZ-type interpolating current and $\Gamma(\Lambda_c\rightarrow
\Lambda\ell^+\nu)=(6.45\pm1.06)\times10^{-14}\mbox{GeV}$ from
Ioffe-type interpolating current. The results show that the adoption
of the Ioffe-type interpolating current is in better agreement with
the experimental data. This is partly due to the fact that in the
case when Ioffe-type interpolator is used, terms proportional to the
mass of the heavy quark appearing in the sum rules (\ref{sr-ioffe})
can play an important role in the numerical analysis. In addition,
the analyses also show that the higher twist contributions are
important for the results so that they need to be included in the
calculation.

\acknowledgments  This work was supported in part by the National Natural Science Foundation of China under Contract No.10675167.


\end{document}